\documentclass{article}
\usepackage[T1]{fontenc}
\usepackage{microtype}
\usepackage{graphicx}
\usepackage{booktabs}
\usepackage{amsmath,amssymb}
\usepackage{xcolor}
\usepackage{xurl}
\usepackage{hyperref}

\usepackage[accepted]{mlsys2025}
\usepackage{algorithm}
\usepackage{algorithmic}
\newcommand{\sys}{LeanSparse}

\mlsystitlerunning{\sys: Memory-Efficient Serving for Native Sparse-Attention LLMs}
\makeatletter
\renewcommand{\Notice@String}{Preprint.}
\makeatother
\hypersetup{pdftitle={LeanSparse: Memory-Efficient Serving for Native Sparse-Attention LLMs},pdfauthor={Zaifeng Pan, Chris Wu, Xinwei Qiang, Zhengding Hu, Zhongkai Yu, Yufei Ding},pdfsubject={Preprint},pdfkeywords={Sparse attention, LLM serving, memory management, KV cache}}
\begin{document}
\twocolumn[
\mlsystitle{\sys: Memory-Efficient Serving for Native Sparse-Attention LLMs}
\mlsyssetsymbol{equal}{*}
\begin{mlsysauthorlist}
\mlsysauthor{Zaifeng Pan}{equal,ucsd}
\mlsysauthor{Chris Wu}{equal,ucsd}
\mlsysauthor{Xinwei Qiang}{ucsd}
\mlsysauthor{Zhengding Hu}{ucsd}
\mlsysauthor{Zhongkai Yu}{ucsd}
\mlsysauthor{Yufei Ding}{ucsd}
\end{mlsysauthorlist}
\mlsysaffiliation{ucsd}{University of California, San Diego, CA, USA}
\mlsyscorrespondingauthor{Zaifeng Pan}{zapan@ucsd.edu}
\mlsyskeywords{Sparse attention, LLM serving, memory management, KV cache}
\vskip 0.3in
\begin{abstract}
Sparse attention limits the history read by each query, but serving long contexts still requires substantial GPU memory. Existing KV offloading systems retain the full history of indexer keys on GPU, limiting decode concurrency. Prefill's temporary memory demand can further delay admission even when enough memory remains for the same request's subsequent decode. We present \sys{}, a serving system that reduces both decode's persistent memory use and prefill's temporary memory demand. Our design exploits two observations: prefill consumes historical data one layer at a time, and decode can identify promising candidates before comparing their original indexer scores. During prefill, \sys{} streams layer histories through two reusable GPU buffers, overlapping next-layer loading with ongoing model computation. During decode, coarse-to-fine indexing searches compact representations and retrieves original indexer keys only for candidate refinement. Separate indexer key and KV caches reuse data across generation steps. We implement \sys{} in SGLang and evaluate it on DeepSeek-V4-Flash. At 950K context length, \sys{} supports 2.95$\times$ as many concurrent requests and achieves 2.14$\times$ decode throughput over HiSparse. On multi-turn Open-SWE traces, it reduces median time to first token from 21.6 to 0.6\,s and median session completion time by 17.7\% relative to ECHO.
\end{abstract}
]
\printAffiliationsAndNotice{\mlsysEqualContribution}
\section{Introduction}
\label{sec:intro}

Native sparse-attention LLMs incorporate sparse selection into training to reduce the cost of processing long contexts~\cite{nsa,moba,deepseekv32,deepseekv4,glm5}. This benefits multi-turn conversations and agent workflows, whose histories accumulate messages, documents, and tool outputs. However, sparse access does not eliminate the need to retain history: tokens skipped by one query may be selected later. The key-value (KV) cache still grows with context length, so concurrent requests can exhaust GPU high-bandwidth memory (HBM) even when each query reads only a small fraction of the stored KV~\cite{echo,hisparse}.

To address this memory pressure, ECHO~\cite{echo} and HiSparse~\cite{hisparse} offload historical KV to host memory and retain a hot KV buffer on GPU. This design exploits the overlap between tokens selected by successive decode steps. Once a token's KV has been fetched, later steps can reuse it from the buffer instead of transferring it again. Prefetching further reduces exposed transfer latency by loading likely selections in advance. These optimizations substantially reduce the GPU memory occupied by attention KV while allowing decoding to continue over a long history.

\begin{figure}[t]
\centering
\includegraphics[width=\columnwidth]{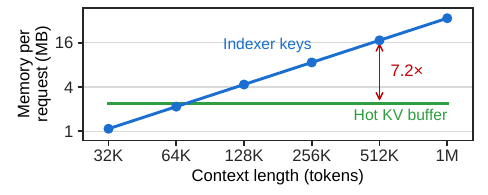}
\caption{Indexer key storage grows with context length. Per-request memory of one CSA layer in HiSparse: FP8 indexer keys versus the fixed hot KV buffer.}
\label{fig:motivation-memory}
\end{figure}

With a learned indexer, however, these systems still retain the full history of \emph{indexer keys} on GPU. The indexer scores these indexer keys to select tokens before attention. They are separate from the KV that attention consumes. Keeping all indexer keys resident allows the indexer to search the history at every decode step, but leaves a memory cost that grows linearly with context length. Figure~\ref{fig:motivation-memory} illustrates this cost for DeepSeek-V4-Flash~\cite{deepseekv4}. At 512K context tokens, its FP8 indexer keys require 7.2$\times$ the space of a 4K-entry hot KV buffer per indexed layer. Thus, reducing the attention-KV cache can leave the indexer as the larger allocation, limiting further concurrency growth.

Besides indexer residency, existing offloading optimizations mainly target decode, where repeated accesses across generation steps make a hot buffer effective. Prefill has a different access pattern. It processes many input queries together, and the tokens they select can span a much larger part of the history than those selected in a single decode step. A newly admitted or resumed request also cannot rely on an already warm GPU buffer. Consequently, prefill cannot benefit from the hot buffer optimization used for decode. HiSparse~\cite{hisparse} thus keeps the full KV history in HBM throughout prefill.

\begin{figure}[t]
\centering
\includegraphics[width=\columnwidth]{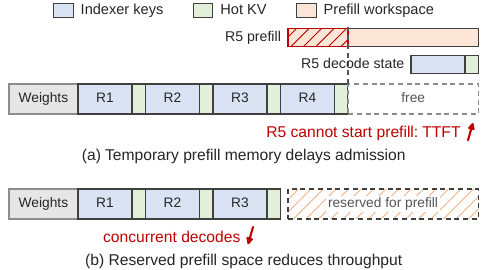}
\caption{Prefill admission competes with decode for GPU memory. A prefill request either waits for decode to release memory or uses a reservation that sits idle when no prefill runs.}
\label{fig:motivation-admission}
\end{figure}

When prefill and decode share a GPU, the remaining HBM may be enough for an arriving request's decode state but insufficient for its temporary prefill demand. As Figure~\ref{fig:motivation-admission}(a) shows, the request must then wait for active decode requests to release memory before prefill can start, increasing time to first token (TTFT). In contrast, reserving HBM for prefill avoids this wait but reduces decode concurrency even when no prefill is running, as shown in Figure~\ref{fig:motivation-admission}(b).

Existing systems~\cite{echo,hisparse} thus suggest prefill-decode (PD) disaggregation~\cite{distserve}, placing the two phases on separate GPU pools. However, instance-level disaggregation duplicates model weights and restricts resource sharing between phases~\cite{expertplex,taichi}. When HBM is already scarce enough to require KV offloading, these extra weight copies further reduce the space available for request state under a fixed GPU budget. Sharing a model instance is therefore a more suitable choice in this setting, as both phases reuse the same weights and memory pool. Efficient memory sharing thus requires reducing both decode's persistent indexer allocation and prefill's temporary memory demand.

To address these limitations, we present \sys{}, a memory-efficient serving system for long-context sparse attention. \sys{} keeps the full history of indexer keys and KV in host memory. It reduces prefill's temporary memory demand through \emph{layer-wise prefill streaming} and decode's persistent indexer residency through \emph{coarse-to-fine indexing}.

\textbf{Layer-wise prefill streaming.} A prefill chunk can extend an already computed prefix, reused from a cached request or produced by earlier chunks of the same request. The prefix's indexer keys and KV have already been computed for each layer, but the new queries consume them one layer at a time. \sys{} streams this prefix state from host memory through two reusable GPU buffers. One serves the current layer while the other loads the next layer's prefix state, without waiting for that layer's new queries. Reusing the buffers reduces prefill's temporary memory demand and leaves more HBM for concurrent decode requests.

\textbf{Coarse-to-fine indexing.} Consecutive decode steps often select overlapping historical positions, yet the indexer still reads every original indexer key to find them. We use a compact approximation to screen the history before retrieving original indexer keys. This first pass retains more candidates than attention needs: positions that are difficult to rank approximately can all survive and be compared using their original scores. Only these candidates require original indexer keys on GPU, and cached indexer keys can serve repeated candidates across steps. Our implementation uses product quantization (PQ)~\cite{pq} for this coarse pass.

We implement \sys{} in SGLang~\cite{sglang} and evaluate it on DeepSeek-V4-Flash~\cite{deepseekv4}. Across context lengths from 200K to 950K tokens, \sys{} supports up to 2.95$\times$ as many concurrent requests and achieves up to 2.14$\times$ the decode throughput of HiSparse~\cite{hisparse}. On multi-turn Open-SWE replay, \sys{} reduces median TTFT by 97.2\% and median session completion time by 17.7\% relative to ECHO~\cite{echo}.

Our contributions are:
\begin{itemize}
    \item We identify two memory bottlenecks that limit long-context sparse-attention serving beyond attention-KV offloading: persistent indexer residency and temporary prefill demand.
    \item We propose \sys{}, a serving system that reduces GPU memory requirements across prefill and decode. It reuses GPU buffers across layers during prefill and retrieves original indexer keys only for candidate refinement during decode.
    \item We implement \sys{} in SGLang and demonstrate higher long-context serving concurrency and throughput, together with lower TTFT and session completion latency on multi-turn workloads.
\end{itemize}

\section{Background}
\label{sec:background}

\begin{figure}[t]
\centering
\includegraphics[width=\columnwidth]{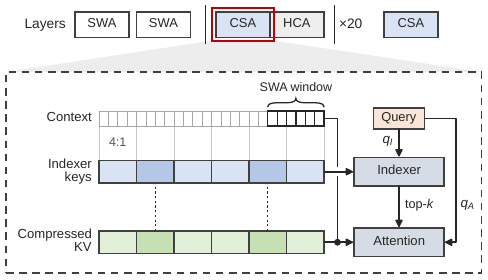}
\caption{DeepSeek-V4-Flash attention architecture and the computation within a CSA layer.}
\label{fig:dsv4}
\end{figure}

\subsection{Sparse Attention in DeepSeek}
Native sparse-attention LLMs reduce the cost of long-context attention by restricting attention computation to a small, query-dependent subset of historical positions~\cite{nsa,moba,deepseekv32,deepseekv4,glm5}. These models are trained with this sparse access pattern so that they can adapt to the selected context and retain modeling quality under a reduced attention budget.

We use DeepSeek-V4-Flash~\cite{deepseekv4} as a concrete example for our serving optimizations. Figure~\ref{fig:dsv4} shows its attention architecture, where compressed sparse attention (CSA) layers alternate with heavily compressed attention (HCA) layers. Each Transformer block also contains a mixture-of-experts (MoE) feed-forward network (FFN). CSA applies DeepSeek Sparse Attention (DSA)~\cite{deepseekv32} to compressed history, using a learned indexer to select historical positions. Both layer types also use sliding-window attention (SWA) for recent tokens. SWA retains a fixed local window, while HCA compresses history at a stride of 128 and attends to all eligible compressed positions. CSA uses a stride of four, retaining far more historical entries while accessing only a small selected subset per query. For long contexts, CSA dominates historical KV storage, and its sparse reads make it suitable for KV offloading. Therefore, both HiSparse~\cite{hisparse} and \sys{} focus their memory optimizations on CSA in this architecture.

Each CSA layer maintains \emph{indexer keys} for selecting historical positions and separate attention KV for computing the output. Figure~\ref{fig:dsv4} illustrates this two-stage computation. For each query, the indexer scores all eligible historical indexer keys and selects up to $k=512$ positions. Core attention then reads the selected compressed KV together with the local SWA window.

\begin{figure}[t]
\centering
\includegraphics[width=\columnwidth]{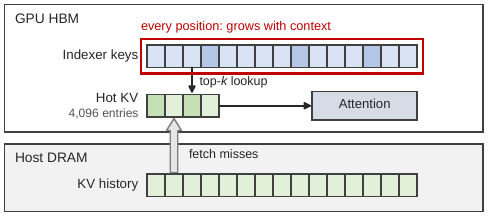}
\caption{HiSparse data placement. Attention KV is offloaded to host memory, while all indexer keys remain on GPU.}
\label{fig:hisparse}
\end{figure}

\begin{figure*}[t]
\centering
\includegraphics[width=\textwidth]{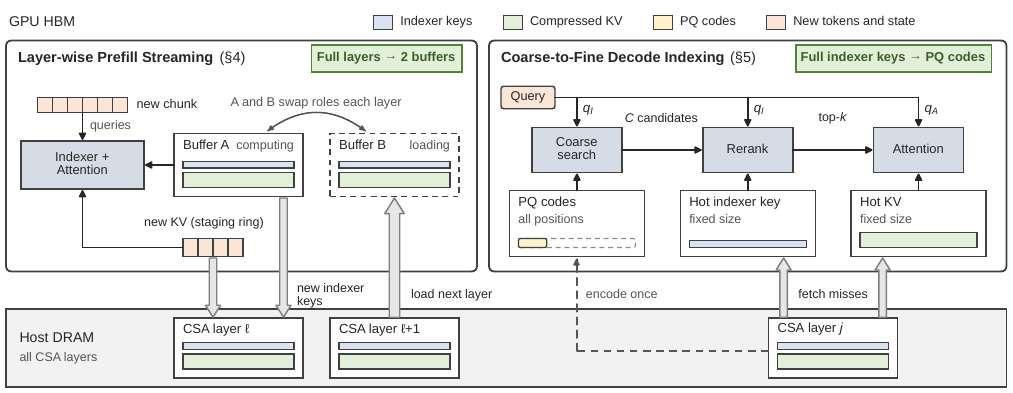}
\caption{System overview of \sys{}. Host memory stores the full history of indexer keys and compressed KV. GPU memory holds two reusable history buffers during prefill, and compact PQ codes with separate indexer key and KV caches during decode.}
\label{fig:architecture}
\end{figure*}

\subsection{Offloading Historical KV for Sparse Attention}

Although sparse attention reduces computation, sparse selection does not directly reduce KV storage. An entry skipped by one query may be selected later, so the complete KV history must be retained. If this history remains in HBM, GPU memory use still grows with context length~\cite{hisparse,echo}.

HiSparse~\cite{hisparse} reduces this GPU memory cost during decode by storing the complete attention-KV history in host memory and maintaining a fixed-capacity hot KV buffer on GPU, as shown in Figure~\ref{fig:hisparse}. At each decode step, the GPU indexer selects Top-$k$ positions. HiSparse reuses selected KV already in the buffer and fetches missing entries from host memory, replacing least recently used entries as needed. Attention then reads the selected KV from the GPU buffer. Because successive decode steps often select overlapping positions, the buffer reduces repeated host transfers while bounding the GPU allocation for attention KV. This design still leaves two memory limitations.

First, HiSparse keeps all historical indexer keys on GPU so that the indexer can score them before each selection. This persistent allocation grows with context length and consumes substantial GPU memory, as shown in Figure~\ref{fig:motivation-memory}. It therefore limits decode concurrency even when the attention-KV allocation is fixed. A straightforward solution would be to offload these indexer keys to host memory as well. However, the GPU indexer must still score all historical indexer keys before it can select the Top-$k$ positions. It would therefore have to fetch the complete set at every decode step, introducing host-to-GPU traffic proportional to context length for every generated token.

Second, HiSparse does not reduce the KV memory required during prefill. It keeps the full attention-KV history in HBM throughout prefill because input queries are processed together and can select KV from a much broader part of the history than a single decode query. A request's temporary prefill demand can therefore exceed its GPU memory footprint after entering decode. When prefill and decode share a GPU, this mismatch can block a new request from starting prefill even though enough HBM remains for its subsequent decode state, as illustrated in Figure~\ref{fig:motivation-admission}(a).

\section{System Overview}
\label{sec:overview}

Figure~\ref{fig:architecture} gives an overview of \sys{}. It addresses these two problems by keeping the full history of indexer keys and compressed KV in host memory and loading data onto GPU differently for prefill and decode. During prefill, each chunk extends a prefix whose indexer keys and compressed KV were computed by a previous request or earlier chunks of the same request. The runtime loads this prefix state one layer at a time and reuses the same GPU buffers across layers. During decode, it searches compact codes on GPU and retrieves original indexer keys only to refine promising candidates. Both paths use the same host copy of the history.

\begin{figure}[t]
\centering
\includegraphics[width=\columnwidth]{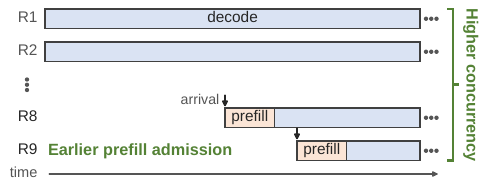}
\caption{Earlier prefill admission and higher decode concurrency in \sys{}.}
\label{fig:serving-workflow}
\end{figure}

As shown in Figure~\ref{fig:serving-workflow}, reusing two GPU buffers reduces prefill's temporary memory demand, allowing new requests to start while others are still decoding. Compact indexer representations reduce decode's persistent memory use per request. Together, these changes support earlier prefill admission and more concurrent decoding under the same GPU memory budget.

We build on SGLang's radix cache~\cite{sglang} to match incoming requests against stored prefixes. The cache maps logical positions to host pages, allowing requests to reuse a prefix without retaining its full GPU footprint. Before each prefill chunk, the runtime constructs indexer key and KV page plans for the already computed prefixes and deduplicates shared compressed KV pages. Logical attention positions are then mapped to their loaded copies or to locally produced state.

Prefill's buffer capacities must accommodate the active batch. The current chunk produces new indexer keys and KV, which remain on GPU until both their consumers and host writeback finish. These entries extend the prefix available to later chunks. Without a cached prefix, the first chunk starts with no prefix state to load.

After prefill, the runtime encodes historical indexer keys and initializes the decode caches. Incomplete pages remain locally accessible as generation extends them. Completed pages join the host-backed history and retain only compact search codes on GPU.

\begin{figure*}[t]
\centering
\includegraphics[width=\textwidth]{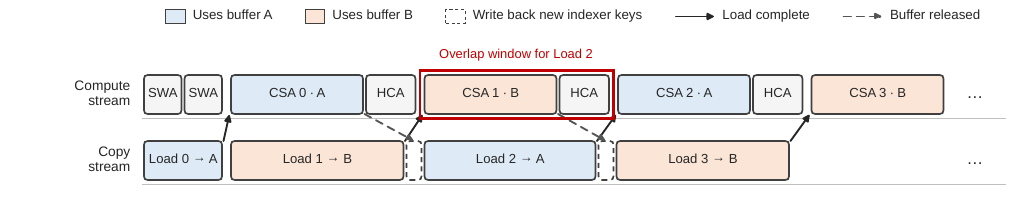}
\caption{Double-buffered prefill execution. Each compute segment includes its block's MoE FFN. The next CSA layer's prefix load overlaps the remaining work in the current CSA block and the intervening HCA block.}
\label{fig:prefill-schedule}
\end{figure*}

The following sections describe layer-wise prefill streaming and coarse-to-fine decode indexing.

\section{Layer-wise Prefill Streaming}
\label{sec:prefill}

To lower prefill's temporary memory demand, \sys{} reuses GPU storage across CSA layers when processing a chunk that extends an already computed prefix. Whether reused from a cached request or produced by earlier chunks of the same request, this prefix has indexer keys and compressed KV already computed for every CSA layer. The current chunk consumes these states one layer at a time, so the GPU copy of a layer's prefix state can be discarded once its consumers finish, while the host copy remains available. After any pending writeback completes, the same GPU storage can hold a later layer's state, reducing temporary prefill demand by avoiding simultaneous residency across all CSA layers.

Because the prefix state has already been computed, it can be loaded before the next layer's queries are available. The indexer needs all prefix indexer keys, while the KV entries selected by attention depend on those queries. \sys{} loads the full compressed KV together with the indexer keys, allowing both transfers to start without waiting for selection, at the cost of moving KV that attention may not use. To overlap these loads with computation, the runtime alternates two GPU buffers: one serves the current layer while the other receives the next layer's prefix state. This pipeline keeps only two layers' loaded prefix state on GPU, and each load serves all queries in the chunk. Section~\ref{sec:prefill-balance} analyzes when computation is long enough to hide the transfer latency. Layer-wise streaming also applies to HCA. For simplicity, our implementation focuses on CSA, which dominates historical state storage.

\subsection{Streaming and Workspace Reuse}

Before processing a chunk, the runtime resolves the prefix's host pages and constructs indexer key and KV page plans for the batch. The compressed KV plan deduplicates pages shared by multiple requests, while the indexer key plan follows request order. Each layer load follows these plans. Without a cached prefix, the first chunk has no prefix state to load.

\paragraph{Overlapping loading and computation.}
The runtime alternates two GPU buffers between computation and loading, as shown in Figure~\ref{fig:prefill-schedule}. While the current CSA layer uses buffer A, a transfer stream loads the next CSA layer's prefix state into buffer B. Computation then switches to B, allowing A to receive the following layer's prefix state once it is safe to reuse.

The runtime starts loading the first CSA layer's prefix state before the model's two initial SWA layers execute. When CSA layer 0 acquires its buffer, the runtime submits the load for CSA layer 1 into B. This transfer can overlap the remaining work in the current Transformer block and the following HCA block, including both MoE FFNs. Each later load follows the same schedule. The compute stream waits for the corresponding load-complete event before consuming a buffer. The transfer stream waits for the buffer's previous consumers and indexer key writeback before overwriting it.

\paragraph{Managing newly produced state.}
Processing the current chunk also produces new indexer keys and compressed KV. New indexer keys are written to host before their GPU buffer is reused. New compressed KV is kept in a separate GPU staging buffer, together with compressed KV from earlier chunks whose host writes are still pending. This buffer is organized as a ring whose slots become reusable only after their GPU consumers and host writeback finish. Attention reads selected KV from either the loaded prefix buffer or this local state. Compression state persists across chunks so that a later chunk can finish a compression window started by the previous one.

As chunks complete and their state is written to host memory, their pages become available for later chunks to load through the same streaming path. At the transition to decode, the runtime builds compact codes from the completed history and retains the recent local tail, as described in Section~\ref{sec:decode}.

\paragraph{Reducing peak memory.}
The runtime allocates both layer buffers at configured capacities. Let $I_{\mathrm{cap}}$ be each buffer's capacity for prefix and newly produced indexer keys, $V_{\mathrm{cap}}$ its capacity for compressed KV loaded from the prefix, and $R_s$ the capacity of the separate staging buffer. The prefill allocation is
\begin{equation}
W_p = 2\left(I_{\mathrm{cap}}+V_{\mathrm{cap}}\right)
 +R_s+W_{\mathrm{act}}+W_{\mathrm{scratch}},
\label{eq:prefill-mem}
\end{equation}
where $W_{\mathrm{act}}$ is the GPU memory occupied by activations for the current chunk, and $W_{\mathrm{scratch}}$ is the temporary workspace used internally by compute kernels while processing that chunk.
Retaining an equally sized buffer for every CSA layer would multiply $I_{\mathrm{cap}}+V_{\mathrm{cap}}$ by the layer count. Streaming keeps this term at two buffers, reducing the HBM reservation needed to admit prefill alongside decode. Actual allocation follows the configured capacities, even when the batch uses fewer pages. Larger prefix batches require larger buffers, and admission must ensure that their page plans fit.

\subsection{Transfer-Compute Balance}
\label{sec:prefill-balance}

Let $\ell$ index successive CSA layers, with an HCA layer between them. On the compute stream, the available overlap window consists of the work between submission of the load for layer $\ell+1$ and the first operation that requires its prefix state. As shown in Figure~\ref{fig:prefill-schedule}, we estimate this window from the remaining work in the current CSA block and the intervening HCA block, including both MoE FFNs. The estimate excludes current-CSA input projections completed before load submission and any preparation of the next CSA layer.

Consider a request appending $B$ tokens to an already computed prefix of length $L$. With CSA compression ratio $r$ and $s_I,s_V$ bytes per indexer key and compressed KV entry, the prefix payload for the next layer is
\begin{equation}
D_{\mathrm{hist}}\simeq \frac{L}{r}(s_I+s_V).
\label{eq:prefill-bytes}
\end{equation}
Let $D_{\mathrm{new}}$ cover additional page traffic serialized with this load, including indexer key writeback. With effective host bandwidth $\beta$ and buffer-reuse and startup delay $\delta$, the time from load submission until the next buffer is ready is
\begin{equation}
T_{\mathrm{load}}\simeq\delta+
\frac{D_{\mathrm{hist}}+D_{\mathrm{new}}}{\beta}.
\label{eq:prefill-transfer}
\end{equation}
Here, bandwidth is the effective rate per GPU under concurrent traffic.

\begin{figure}[t]
\centering
\includegraphics[width=\columnwidth]{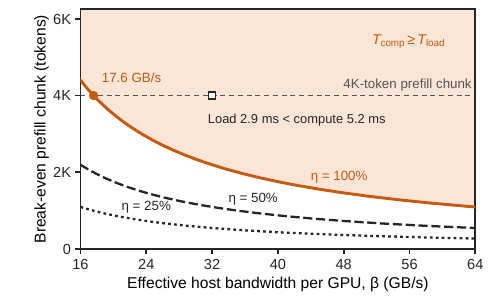}
\caption{Modeled prefill chunk needed to cover the next-layer transfer for DeepSeek-V4-Flash with a 512K-token prefix on H100 SXM with TP4. The window includes both blocks' MoE FFNs. Curves vary compute efficiency $\eta$ with startup and reuse delay $\delta=0$.}
\label{fig:prefill-roofline}
\end{figure}

We apply the roofline model~\cite{roofline} to the serial operation groups in this window. Let $F_g$ and $Q_g$ be a group's FLOPs and HBM traffic, $P_g$ its compute ceiling, $\eta$ a common compute-efficiency factor, and $\gamma$ the HBM bandwidth ceiling. Then
\begin{equation}
T_{\mathrm{comp}}(\eta)\simeq
\sum_{g\in\mathcal{G}_{\mathrm{CSA}}\cup\mathcal{G}_{\mathrm{HCA}}}
\max\!\left(\frac{F_g}{\eta P_g},\frac{Q_g}{\gamma}\right).
\label{eq:prefill-compute}
\end{equation}
$\mathcal{G}_{\mathrm{CSA}}$ includes indexing, sparse attention, output projections, and the current block's MoE FFN after load submission. $\mathcal{G}_{\mathrm{HCA}}$ includes the intervening block's input and output projections, HCA attention, and MoE FFN. Counts follow the model dimensions and tensor-parallel placement. HBM traffic assumes ideal KV and weight reuse and a minimum seven-expert working set per block. Routing, selection, compression, normalization, launch, and collective overheads are omitted.

Under these assumptions, the predicted wait for the next buffer is
\begin{equation}
T_{\mathrm{wait}}=\max\!\left(0,T_{\mathrm{load}}-T_{\mathrm{comp}}(\eta)\right).
\label{eq:prefill-threshold}
\end{equation}
Larger chunks increase the work sharing each prefix load, while the dominant prefix payload stays fixed.

Figure~\ref{fig:prefill-roofline} plots the modeled threshold with $\delta=0$ and dense FP8 compute ceilings. At $\eta=1$, a 4K-token chunk provides a 5.2\,ms window, including 0.36\,ms from the two MoE FFNs. Loading and serialized writeback take 2.9\,ms at 32\,GB/s effective host bandwidth per GPU, leaving a modeled startup and reuse delay budget of about 2.4\,ms. This margin depends on the effective transfer bandwidth under concurrent computation.

With multiple requests, computation over the prefixes scales with $\sum_j B_jL_j$, while loading scales with $\sum_j L_j$ for disjoint prefixes. Total batch size alone therefore does not determine overlap. The first load has a different window because it overlaps the initial two Transformer blocks, each containing SWA and an MoE FFN.

\section{Coarse-to-Fine Decode Indexing}
\label{sec:decode}

HiSparse~\cite{hisparse} observes that consecutive decode steps select many of the same historical positions. A small GPU cache can therefore serve much of the attention demand. The indexer, however, still reads every historical indexer key to find the next Top-$k$. Keeping these indexer keys on GPU consumes memory proportional to context length. Moving them to host would instead require a full-history transfer at every step.

This repeated selection of historical positions across decode steps suggests that, over a short decode interval, attention repeatedly returns to a limited part of the history. This motivates using a compact approximation to identify likely candidates before reading their original indexer keys. Crucially, this approximation only needs to retain a useful candidate set. Even if it cannot distinguish two positions near the Top-$k$ boundary, it can keep both and leave that decision to a second pass using their original scores.

\sys{} applies this coarse-to-fine approach to indexer key offloading. It searches compact representations of the full history, retrieves original indexer keys for the surviving candidates, and reranks them to select the final Top-$k$. Retrieved indexer keys are cached for later steps. The full history remains searchable, while only a small set of original indexer keys needs to reside on GPU or cross the host link at each step.

At each decode step, the indexer searches compact representations of the history and retains $C>k$ candidates. The extra candidates allow positions displaced by approximation error to remain available for refinement. Because this search covers the full history, it can discover positions that were neither selected nor cached at the previous step.

\begin{figure}[t]
\centering
\includegraphics[width=\columnwidth]{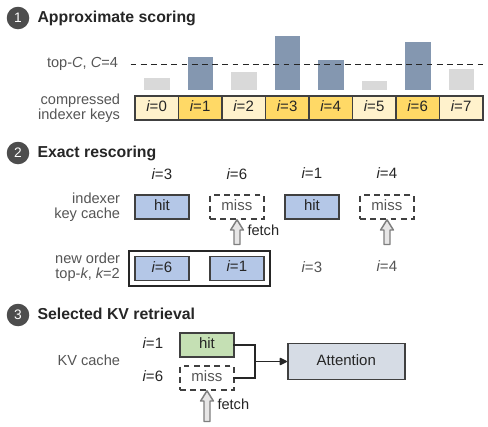}
\caption{Coarse-to-fine indexing with $C=4$ and $k=2$. Refinement changes the candidate ranking, so the indexer key cache and KV cache serve different positions.}
\label{fig:decode-example}
\end{figure}

\paragraph{Approximate scoring with PQ.}
Our implementation uses product quantization (PQ)~\cite{pq} for coarse scoring. We split each 128-dimensional indexer key into 32 four-dimensional subvectors and encode each with a one-byte index into a layer-specific codebook of 256 centroids trained offline. The resulting PQ32 code occupies 32 bytes, compared with 132 bytes for the original FP8 indexer key and its scale. The original representation remains in host memory for refinement.

The coarse scorer substitutes codebook centroids for the original subvectors and applies the indexer's weighted head aggregation:
\begin{equation}
\widetilde{s}_i=\sum_h w_h\,\mathrm{ReLU}\!\left(\sum_j
q_{h,j}^{\mathsf T}\mathcal{C}_j[c_{i,j}]\right),
\label{eq:pq-score}
\end{equation}
where $q_{h,j}$ is head $h$'s query subvector, $w_h$ its weight, and $\mathcal{C}_j[c_{i,j}]$ the centroid selected by indexer key $i$ in subspace $j$. As generation extends the history, the runtime keeps recent indexer keys in a GPU tail and encodes each completed page when writing it to host, allowing the tail storage to be reused.

\paragraph{Retrieving and refining candidates.}
Consecutive decode steps often select many of the same historical positions, allowing refinement to reuse the corresponding indexer keys from earlier steps. Keeping these indexer keys in a GPU cache reduces host reads to candidates that are not already cached. This reuse also extends to positions that repeatedly enter the candidate set without surviving the final Top-$k$, whose indexer keys remain useful even when their KV is not needed. \sys{} therefore maintains separate fixed-capacity caches for original indexer keys and compressed KV.

Before fetching original indexer keys, \sys{} retains all candidates that hit in the indexer key cache and filters low-scoring misses. In the evaluated configuration, a miss is discarded if its approximate score is below the candidates' $k$th-largest score minus one sample standard deviation. Only the remaining misses require host transfers.

\sys{} computes original indexer scores for the retained candidates and the recent indexer keys in the GPU tail. It selects the final Top-$k$ and retrieves their compressed KV from the KV cache or host memory for attention. Figure~\ref{fig:decode-example} shows how refinement changes the approximate ranking.

\begin{figure*}[t]
    \centering
    \includegraphics[width=\textwidth]{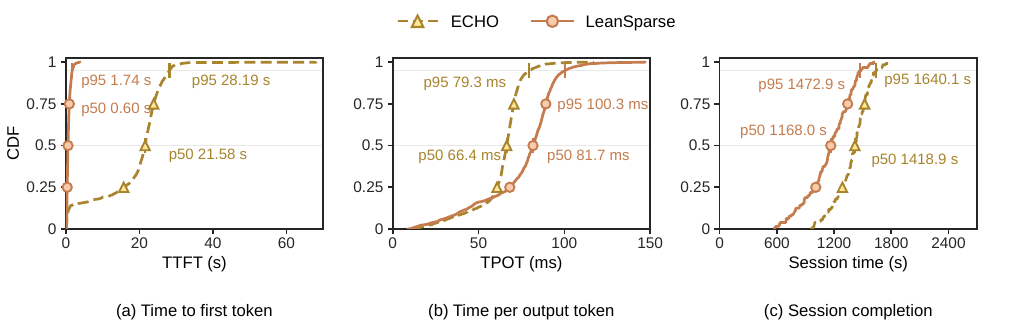}
    \caption{Open-SWE replay latency distributions: (a) TTFT and (b) TPOT for the same 2,432 continuation calls, and (c) completion time for 128 full sessions, including tool waits. Vertical ticks and labels indicate p50 and p95.}
    \label{fig:eval-replay}
\end{figure*}

Refinement cannot recover positions removed by coarse selection or miss filtering, but we observe that it recovers nearly all positions selected by the original indexer and achieves substantially higher token-selection accuracy than selecting Top-$k$ directly from PQ32 scores. Section~\ref{sec:eval-accuracy} reports the selection accuracy and downstream model quality.

\paragraph{Overlapping transfers with computation.}
Indexer key retrieval must finish before refinement can determine the KV to retrieve. To overlap these transfers with computation, \sys{} divides a decode batch into request groups. One group's indexer key reads can overlap another group's coarse scoring or refinement, while its KV reads can overlap later groups' refinement.

\section{Evaluation}
\label{sec:eval}

We evaluate how \sys{}'s lower GPU memory use translates into serving performance. Multi-turn replay measures response latency and session completion time when prefill and decode share GPU resources. Long-context decode experiments examine concurrency and throughput as context length grows. Finally, we evaluate model accuracy on reasoning, long-context understanding, and retrieval tasks to assess the effect of approximate candidate filtering.

\subsection{Experimental Setup}

We run DeepSeek-V4-Flash~\cite{deepseekv4} with four NVIDIA H100 GPUs and tensor parallelism in SGLang~\cite{sglang}. Multi-turn serving uses our implementation of ECHO~\cite{echo} because its open-source release does not support DeepSeek-V4. HiSparse~\cite{hisparse} supports only disaggregated prefill and decode, so we exclude it from this experiment. Since the two baselines have nearly identical decode paths, decode-only experiments compare against HiSparse.

\sys{} uses 32-byte PQ codes and refines 2K candidates to select 512 entries. Its per-request GPU caches hold 4K indexer keys and 2K KV entries per CSA layer. HiSparse uses a 4K-entry KV buffer. ECHO shares 256K KV entries per layer and uses 512-entry per-request buffers. Replay enables prefix caching and mixed prefill-decode batches, with 4K-token prefill chunks and up to 128 active requests.

\subsection{Multi-Turn Serving Performance}

\paragraph{Workload.}
We replay NVIDIA Open-SWE-Traces~\cite{openswetraces} using 128 long-context sessions, excluding codebook-training sessions. Each contributes a 20-call window ending at its longest prompt, preceded by a warmup prefill outside the measured interval. Sessions arrive as a Poisson process at 120 per minute. Calls remain sequential, with tool waits reconstructed from reported durations and command-based estimates. We replay recorded prompts and output lengths, capped at 8K output tokens and the context limit. Later prompts retain their recorded contents.

TTFT and TPOT statistics cover the same 2,432 continuation calls after each session's first measured call. Time per output token (TPOT) is the first-to-last-token interval divided by the number of inter-token gaps, excluding calls with fewer than two output tokens. Session time spans the first measured submission through the final completion, including tool waits. Throughput uses all 2,560 measured calls, dividing their total output tokens by the full replay duration.

\paragraph{Results.}
\sys{} substantially reduces response startup delays. Figure~\ref{fig:eval-replay}(a) shows median TTFT falling from 21.6 to 0.6\,s, while the 95th percentile falls from 28.19 to 1.74\,s. Shorter waits improve responsiveness across the measured call distribution, including slow calls.

These shorter waits also help complete sessions sooner. Figure~\ref{fig:eval-replay}(c) shows median session completion time decreasing by 17.7\%. Both methods complete all calls. Because calls are sequential, faster response starts help sessions advance. Token generation and tool waits still contribute to completion time, so the session-level reduction is smaller than the TTFT reduction. Session time captures how these delays accumulate across dependent calls.

Figure~\ref{fig:eval-replay}(b) shows median TPOT increasing from 66.4 to 81.7\,ms. We attribute this primarily to larger decode batches and earlier prefill scheduling after memory pressure is reduced. More active requests and prefill work share each execution step, increasing the interval between output tokens. Sessions still finish sooner, and aggregate throughput rises by 10.2\%.

\paragraph{Indexer selection recall versus direct quantization.}
In an offline analysis of Open-SWE traces, we compare our coarse-to-fine approach with direct index scoring using PQ32. Recall measures the fraction of the exact indexer's top-512 entries recovered by the selected set. Selecting 512 entries directly from PQ32 scores achieves 77.41\% mean recall. Expanding the candidate set to 2,048 and reranking with exact scores raises mean recall to 99.49\%. These results show that coarse-to-fine indexing combines compact GPU-resident representations with high selection recall.

\subsection{Long-Context Decode Performance}

\begin{figure}[!t]
    \centering
    \includegraphics[width=\columnwidth]{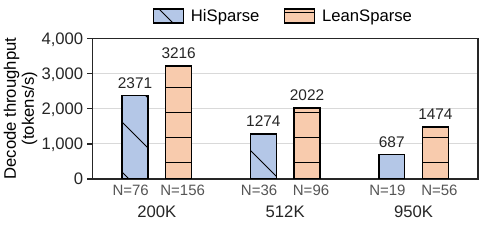}
    \caption{Decode throughput at measured capacities. N denotes concurrency.}
    \label{fig:eval-decode}
\end{figure}

\begin{figure}[!t]
    \centering
    \includegraphics[width=\columnwidth]{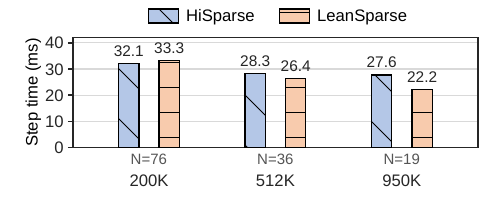}
    \caption{Decode step time at equal concurrency. N denotes concurrency.}
    \label{fig:eval-decode-cost}
\end{figure}

\paragraph{Workload.}
We construct 200K-, 512K-, and 950K-token contexts from LongSWE-Bench~\cite{longcodebench}, trimming repository files while preserving instructions and issue descriptions. Prefix reuse is disabled for these requests. All admitted requests finish prefill before decode begins, with outputs capped at 8K tokens. After a 60-second settling interval, we measure steady decode over 2K new tokens per stream. We report steady-state throughput at the maximum concurrency reached in each stress test. For the step-time comparison, both methods use HiSparse's maximum tested concurrency at each context length.

\paragraph{Results.}
\sys{} admits 2.05-2.95$\times$ as many concurrent requests across the three context lengths. Figure~\ref{fig:eval-decode} shows that its throughput advantage grows from 1.36$\times$ at 200K to 2.14$\times$ at 950K. Compact indexer representations leave more GPU space for concurrent requests as histories grow. The throughput improvement is smaller than the concurrency increase because admitting more streams also lengthens each decode step.

Figure~\ref{fig:eval-decode-cost} compares decode step time at equal concurrency. \sys{} takes 3.8\% longer per step at 200K but 19.6\% less time at 950K. At shorter contexts, savings in exact indexer scoring are small relative to the added coarse-screening overhead. Longer histories increase HiSparse's exact scoring work, while \sys{} restricts exact scoring to 2K candidates after approximate filtering. The coarse pass still scans the compact history, so its cost grows with context length. The candidate budget bounds the subsequent reads of original indexer keys and exact scoring.

Both systems admit fewer requests as context length grows. \sys{} reduces the storage per historical position, but its compact representations still grow with the history. Its capacity advantage persists across the tested lengths, allowing more requests to share the GPU even as longer contexts increase each request's memory demand.

\subsection{Model Accuracy}
\label{sec:eval-accuracy}

\paragraph{Tasks.}
We compare \sys{} with unmodified SGLang and HiSparse on GSM8K~\cite{gsm8k}, multi-key needle-in-a-haystack (NIAH)~\cite{ruler}, LongBench v2~\cite{longbenchv2}, and two-needle OpenAI MRCR~\cite{mrcr}. GSM8K uses 20 fixed demonstrations and all 1,319 test questions. NIAH contains 150 prompts at 32K, 128K, and 200K. LongBench v2 includes 403 questions fitting within 245K tokens, and MRCR contains 265 examples. For the comparison in Table~\ref{tab:accuracy}, all methods use greedy non-thinking decoding, a 256K-token context limit, 4K-token prefill chunks, and up to 64 concurrent requests. Scores measure answer accuracy, recovered-needle fraction, or MRCR reference-text similarity.

\paragraph{Results.}
Table~\ref{tab:accuracy} shows that \sys{} largely preserves model quality across the tested tasks, with all scores within 0.5 percentage points of SGLang. NIAH recovery remains complete, and the other tasks show small changes in both directions. Paired McNemar tests find no significant difference on GSM8K or LongBench v2. This is consistent with our coarse-to-fine design, which uses approximate scores only to narrow the candidate set and computes exact scores from the original indexer keys for final selection. The results suggest that this filtering step reduces memory use without materially affecting accuracy on these workloads.

\begin{table}[!t]
    \centering
    \caption{Model quality with a 256K context limit. Scores are percentages. Higher is better.}
    \label{tab:accuracy}
    \begin{tabular}{lrrrr}
        \toprule
        Method & GSM8K & NIAH & LBv2 & MRCR \\
        \midrule
        SGLang & 95.38 & 100.00 & 50.87 & 90.19 \\
        HiSparse & 95.30 & 96.00 & 51.61 & 87.86 \\
        \sys{} & 95.83 & 100.00 & 50.62 & 90.62 \\
        \bottomrule
    \end{tabular}
\end{table}

\section{Related Work}
\label{sec:related}

\paragraph{LLM serving.}
PagedAttention~\cite{pagedattention} and SGLang~\cite{sglang} improve KV allocation and prefix reuse. Sarathi-Serve~\cite{sarathi} and DistServe~\cite{distserve} address prefill-decode interference through chunked scheduling and phase separation. FlexGen~\cite{flexgen} distributes model state across GPU, CPU, and disk for throughput-oriented inference. CachedAttention~\cite{cachedattention} combines hierarchical KV reuse with layer-wise preloading, while Mooncake~\cite{mooncake} extends KV reuse across distributed caches. For selective offloading, InfiniGen~\cite{infinigen} predicts next-layer KV demand, while FreeKV~\cite{freekv} combines speculative retrieval with correction and streamed transfers. \sys{} reduces persistent decode state and temporary prefill storage for active requests.

\paragraph{Sparse attention.}
Training-free methods reduce attention work without retraining: MInference~\cite{minference} exploits sparse prefill patterns, and Quest~\cite{quest} selects query-dependent KV pages. ArkVale~\cite{arkvale}, PQCache~\cite{pqcache}, and ShadowKV~\cite{shadowkv} use compact representations to retrieve offloaded KV selectively. Native Sparse Attention~\cite{nsa}, MoBA~\cite{moba}, and DeepSeek Sparse Attention~\cite{deepseekv32} incorporate sparse selection into training. HISA~\cite{hisa} and DoSA~\cite{dosa} accelerate indexing through candidate filtering. SparseServe~\cite{sparseserve} combines KV offloading and batch control with layer-segmented prefill to reduce prefill KV residency, but does not discuss loading cached prefixes from host memory or overlapping these loads with prefill computation. ECHO~\cite{echo} and HiSparse~\cite{hisparse} manage KV residency through caching and prefetching. \sys{} combines coarse-to-fine retrieval of separate indexer keys with double-buffered prefill streaming.

\section{Conclusion}
\label{sec:conclusion}
Native sparse-attention serving remains constrained by full indexer histories on GPU and temporary prefill demand. These costs limit decode concurrency and delay prefill admission. We present \sys{} to address both costs. Its prefill path streams layer histories through two reusable buffers, while coarse-to-fine decode indexing retrieves original indexer keys only for candidate refinement. Our evaluation shows that \sys{} achieves higher decode throughput, shorter response delays, and faster session completion than state-of-the-art baselines.

\label{page:lastmain}
\raggedbottom
\bibliographystyle{mlsys2025}
\bibliography{references}

\section*{AI Usage Disclosure}
OpenAI Codex assisted with writing and revision and literature search. Anthropic Claude assisted with code implementation, experiments, and figure preparation. The authors take full responsibility for the correctness, originality, and integrity of the paper.
\end{document}